\documentstyle[11pt,nyas]{article}
\begin{document}

\title{An attempt to do without dark matter}
\author{William H. Kinney and Martina Brisudova}
\affil{University of Florida, Gainesville}

\begin{abstract}
The discrepancy between dynamical mass measures of objects such as galaxies and the observed distribution of luminous matter in the universe is typically explained by invoking an unseen ``dark matter'' component. Dark matter must necessarily be non-baryonic. We introduce a simple hypothesis to do away with the necessity for dark matter by introducing an additional non-gravitational force coupled to baryon number as a charge. We compare this hypothesis to Milgrom's Modified Newtonian Dynamics. The model ultimately fails when confronted with observation, but it fails in an interesting way.
\end{abstract}

\section{Introduction}

In a dynamical sense, the universe does not behave the way we would expect based on the observed distribution of matter. Virtually every object at galactic scales and larger is being pulled on harder than can be explained simply by gravity and the distribution of matter we can see directly. The conventional (and very successful) explanation for this inconsistency is that there is matter we can't see, {\em dark matter}, that is responsible for the additional gravitational force. There is, however, an obvious alternative to this explanation. It could be that something about gravity changes on very large scales and is responsible for the observed dynamics of galaxies and clusters of galaxies. In fact, it would be quite surprising if general relativity, which is fairly well tested on solar system scales, were to survive unmodified when one considers scales relevant to cosmology, an extrapolation of some 14 orders of magnitude! While the standard cosmology is elegant and powerful, it is instructive (and entertaining) to consider alternatives. In this paper we describe one such alternative model. This model ultimately fails when faced with observational data, but it fails in a sufficiently interesting way that it is worth the time spent considering it. 

It is not clear how far one has to go in order to construct such an alternative. Is it possible to save a reasonably ``standard'' cosmology in the absence of dark matter, i.e. an expanding spacetime arising out of a hot big bang with a horizon and a uniform Hubble expansion? Is it possible to leave the framework of general relativity intact? Is it possible even to retain Newton's laws? One famous attempt to construct a cosmology with no dark matter is Milgrom's {\em Modified Newtonian Dynamics} (MOND),$^1$ which postulates that Newton's law of gravity fails at large distances. MOND introduces a fundamental acceleration $a_0$, such that the ``true'' acceleration of a body is related to the Newtonian acceleration $\vec a_N$ by an interpolating function $\mu$:
\begin{equation}
\vec a_N = \mu\left(a / a_0\right) \vec a,
\end{equation}
where $\mu\left(x\right)$ is taken to have the behavior
\begin{equation}
\mu\left(x\right) \longrightarrow \Bigg\lbrace {x\ \left(x \ll 1\right) \atop 1\ \left(x \gg 1\right)}.
\end{equation}
This hypothesis easily explains the most compelling observational evidence for the existence of dark matter, which is the flatness of galactic rotation curves. It is a simple matter to see than the MOND hypothesis gives flat rotation curves as a natural consequence. Taking the acceleration of a body in a circular orbit of radius $r$,
\begin{equation}
a = \left({v^2 \over r}\right),
\end{equation}
we write the relation to the Newtonian gravitational force in the MOND limit $\mu\left(x\right) \rightarrow x$ as
\begin{equation}
a_N \rightarrow {a^2 \over a_0} = {v^4 \over a_o r^2} = {G M \over r^2}.
\end{equation}
The orbital velocity $v$ is then just a combination of constants:
\begin{equation}
v^4 = G M a_0 = {\rm const.},
\end{equation}
so that the observed flat rotation curves of galaxies at large radii are a natural consequence of MOND. The constant $a_0$ can be determined from rotation curve data, with the ``standard'' value being
\begin{equation}
a_0 = 1.1 \times 10^{-10} h_{70}\ {\rm m/s^2},
\end{equation}
where $h_{70}$ is the value of the Hubble constant $H_0$ in units of $70\, {\rm km / sec / Mpc}$. MOND has in fact been shown to work very well as an explanation for the observed rotation curves of galaxies.$^{2,3}$ It is, however, a radical hypothesis. Consider the force between two objects of mass $m$ and mass $M$, respectively, in the MOND limit:
\begin{eqnarray}
F_{\rm MOND} =&& m \sqrt{a_0 a_N}\cr
\propto&& m \sqrt{M}.
\end{eqnarray}
We see that one must abandon even Newton's rule that every action has an equal and opposite reaction. Surely this is too high a price to pay for explaining even a large body of astrophysical data! We wish to construct a less sweeping alternative, one which leaves as much standard physics intact as possible: we wish to get rid of the bath water, but still keep the baby.

\section{A more palatable alternative to MOND?}

In this section we construct an alternative to MOND which provides a good fit to cosmological observations without the requirement either for dark matter or for radical revisions to basic physics. As discussed in the previous section, this model is motivated by the observation that the standard cosmology requires gravity to be the only relevant force operating over a span of 14 or more orders of magnitude in scale, which would make it unique among fundamental forces. It is certainly a possibility that a new, non-gravitational force becomes dominant on very large scales,
\begin{equation}
V\left(r\right) = - {G M m \over r} + V_1\left(r\right),
\end{equation}
where the potential due to the non-gravitational component is labeled $V_1$. Such a new interaction must couple to a charge of some kind, and we note that standard particle physics provides an excellent candidate: baryon number. An interaction mediated, for example, by a scalar particle coupled to baryon number would act as a universal attractive force similar to gravity, but with distinct scales and couplings. A naive model of a scalar-mediated force, however, will not work for our purpose since by Gauss' Law, the force must fall of as $1 / r^2$ and if the new force is to dominate over gravity at large scales it must do so at {\em all} scales. We postulate instead a force which falls off as $1 / r$, with potential
\begin{equation}
V_1\left(r\right) = \alpha \ln\left(r\right),
\end{equation}
where the constant $\alpha$ is given in terms of the baryon numbers $b$ of the interacting bodies and a fundamental mass scale $\Lambda$,
\begin{equation}
\alpha = \Lambda b_1 b_2.
\end{equation}
We have no fundamental model which creates this behavior, but simply take it as a phenomenological guess. Such a logarithmic potential leads naturally to flat rotation curves. The acceleration of a body in orbit around a central mass $M$ can be written
\begin{eqnarray}
a\left(r\right) =&& {G M \over r^2} + {\alpha \over m r}\cr
=&& {\alpha \over m r} \left[1 + \left({r_0 \over r}\right)\right],
\end{eqnarray}
where the fundamental length scale $r_0$ is defined as
\begin{equation}
r_0 \equiv {G M m \over \alpha}.
\end{equation}
Then, for $r \gg r_0$,
\begin{equation}
a \rightarrow {\alpha \over m r} = {v^2 \over r},
\end{equation}
and we have flat rotation curves
\begin{equation}
v^2 = {\alpha \over m} = {\rm const.}
\end{equation}
To be precise, the rotation velocity is constant as long as the baryon number to mass ratio of the galactic matter is constant, since
\begin{eqnarray}
\alpha =&& \Lambda b_1 b_2 = \Lambda \left({b_1 \over M}\right) \left({b_2 \over m}\right) M m\cr
\simeq&& {\Lambda M m \over m_p^2},
\end{eqnarray}
where $m_p$ is the proton mass. This expression is valid up to the proton/neutron mass ratio. The rotation velocity of a body about a galaxy can then be expressed as a ratio of mass scales,
\begin{equation}
v^2 \simeq {\Lambda M_{\rm gal} \over m_p^2}.
\end{equation}
This model can be cast in a form very similar to MOND, in which the true acceleration and the Newtonian acceleration are related through an interpolating function,
\begin{eqnarray}
a_N =&& {1 \over 1 + \left(r / r_0\right)} a\cr
=&& \mu\left(a / a_0\right) a,
\end{eqnarray}
where the asymptotic acceleration $a_0$ is given by
\begin{equation}
a_0 = {\alpha \over m r_0},
\end{equation}
and the interpolating function $\mu\left(x\right)$ is given by
\begin{equation}
\mu\left(x\right) = {\sqrt{1 + 4 x} - 1 \over \sqrt{1 + 4 x} + 1}.
\end{equation}
It is important to note, however, that this model is {\em not} equivalent to MOND, since the asymptotic acceleration $a_0$ is not a fundamental constant, but varies from galaxy to galaxy:
\begin{equation}
a_0 = {\alpha \over m r_0} \simeq \left({\Lambda^2 \over G m_p^2}\right) M_{\rm gal}.
\end{equation}
Instead, the radius $r_0$ is a universal constant,
\begin{equation}
r_0 = {G M m \over \alpha} \simeq {G m_p^2 \over \Lambda}.
\end{equation}
To explain galactic rotation curves, the fundamental radius $r_0$ must be of order $10\,{\rm kpc}$ or so. (In the next section we derive an independent estimate of $r_0$ based on X-ray observations of galaxy clusters.) Unlike the breakdown of Newtonian physics which occurs in MOND, $F_{\rm MOND} \propto m \sqrt{M}$, Newton's third law is preserved,
\begin{equation}
F \propto {b_1 b_2 \over r} \sim {M m \over r}.
\end{equation}
Our hypothesis leaves virtually all of standard physics intact. 

Flat rotation curves, however, are but one class of a host of astrophysical observations which must be fit with such a model. In the next section, we describe the confrontation of our hypothesis with observation. Remarkably, the model fits a variety of independent constraints, although we ultimately find the model lacking and reject it as a plausible scenario.

\section{Confrontation with observation}

Galactic rotation curves are just the beginning. Our model provides a tidy, minimalist explanation for flat rotation curves without the requirement for non-baryonic dark matter. Such an additional force, however, will dominate over gravity on all scales $r > r_0$ and will therefore be subject to a large number of observational constraints over a range of scales, including galactic scales, cluster scales, large-scale structure, and the universe as a whole:
\begin{list}{$\bullet$}{}
\item Galaxies
\begin{list}{-}{}
\item Galactic masses
\item Tully-Fisher relation
\end{list}
\item Clusters
\begin{list}{-}{}
\item Dynamical mass measures
\item X-ray mass measures
\item Lensing
\end{list}
\item Large Scale Structure
\begin{list}{-}{}
\item Silk damping and primordial perturbations
\end{list}
\item Cosmology
\end{list}
We include cosmology in the list because general relativity is left intact in our model, and the Friedmann-Robertson-Walker universe is still a viable cosmological model. The extra force simply adds a term to the stress-energy in the Einstein field equations:
\begin{equation}
G_{\mu \nu} = 8 \pi G \left[T^0_{\mu \nu} + \delta T_{\mu \nu}\right],
\end{equation}
where $T^0_{\mu \nu}$ is the usual stress-energy of the baryons, and $\delta T_{\mu nu}$ is the contribution from the interaction. A fully relativistic model is required to calculate $\delta T_{\mu \nu}$, and we do not consider this question further here. Large scale structure provides a strong constraint on a baryon-only cosmology as well, since photon diffusion (Silk damping) erases primordial fluctuations in the baryon fluid on scales smaller than the horizon at early times. We also leave this important question hanging, and instead concentrate on galactic and cluster scales. We find that these smaller scales alone are sufficient to rule out the model. 

We first consider galactic scales. The first and simplest question we can ask is what is the typical mass of a galaxy? Note that we are assuming that galaxies are made up entirely of baryons, so galactic masses should be neither too large nor too small relative to their luminosities. We can write the mass of the galaxy in terms of the rotation velocity $v_c$ as
\begin{eqnarray}
M_{\rm gal} =&& {r_0 v_c^2 \over G}\cr
=&& 1.1 \times 10^{10} M_{\odot} \left({r_0 \over 1\,{\rm kpc}}\right) \left({v_c \over 220\,{\rm km/s}}\right)^2.
\end{eqnarray}
This figure is nicely consistent with typical galaxy luminosities of ${\cal L} \sim {\rm few} \times 10^{10} {\cal L}_{\odot}$, so the assumption of a baryon-only halo works very well in this model. We can gain further constraints on galactic mass-to-light ratios by use of the Tully-Fisher relation, an empirical relation between a galaxy's luminosity and rotation velocity,
\begin{equation}
{\cal L} \propto v_c^4.
\end{equation}
The simplest and most physical assumption for a baryon-only halo is that a galaxy's luminosity is somewhere close to proportional to its mass, ${\cal L} \propto M_{\rm gal}$. That is, all baryons are roughly equally luminous. This is in fact a natural consequence of MOND, since MOND predicts $M_{\rm gal} \propto v_c^4$ and therefore $M_{\rm gal} \propto {\cal L}_{\rm gal}$ from the Tully-Fisher relation. Explaining flat rotation curves with an extra force, however, requires a strong variation of luminosity with mass, since $M_{\rm gal} \propto v_c^2$, and, from Tully-Fisher,
\begin{equation}
{\cal L}_{\rm gal} \propto v_c^4 \propto M_{\rm gal}^2.
\end{equation}
While it is difficult to understand why a galaxy might have such a relationship between mass and luminosity, neither is it ruled out by observation. We will simply take it as a prediction of the model. 

What about larger scales? The rich data available on galaxy clusters provide several useful constraints on the model. We can divide observations of galaxy clusters into three general classes. First, observations at X-ray wavelengths and measurements of the Sunyaev-Zel'dovich (SZ) effect in clusters provide a direct view of the baryons in the cluster. They allow us to ``count'' the baryons directly. Dynamical measures such as velocity dispersion or mass measurements based on assumptions of hydrostatic equilibrium probe the binding energy of the cluster. With the assumption of Newtonian gravity, these measures actually ``weigh'' the cluster, but with the assumption that a non-gravitational force dominates on cluster scales, dynamical measurements in fact probe the form of the force. We show below that combining ``baryon counting'' measurements and dynamical measurements allows us to fix $r_0$. The third class of measurements is gravitational lensing, which we assume (unlike the dynamical measurements) to be governed by General Relativity and produce a direct measure of the mass of the cluster. We save a discussion of lensing for last, because it rules out the model. 

With the assumption that Newtonian gravitation is the dominant force binding a galaxy cluster together, a combination of X-ray measurements and dynamical measurements is generally used to fix the ``baryon fraction'' $f_{\rm B}$ of the cluster. We will assume a typical value for $f_{\rm B}$ of$^4$
\begin{equation}
f_{\rm B} = 0.06 h^{-3/2}
\end{equation}
In the standard picture, such a determination of the cluster baryon fraction is powerful evidence for the existence of dark matter. Since we wish to do without dark matter, we shall assume that the {\em true} baryon fraction is unity $f_{\rm B} = 1$ and that they apparent baryon fraction of galaxy clusters is an artifact of the additional force which dominates over gravity at cluster scales. This is sufficient to fix the fundamental length $r_0$. Consider a cluster with velocity dispersion $\sigma$ and core radius and density $r_c$ and $\rho_c$. In a simple model of such a cluster,$^5$ the density as a function of radius can be written
\begin{equation}
a \rho\left(r\right) = {3 \rho_c \sigma^2 r \over r_c^2},
\end{equation}
where $a$ is the acceleration. In our model of an extra force, the true acceleration is related to the Newtonian acceleration $a_N$ as
\begin{equation}
a = \left({r \over r_0}\right) a_N,
\end{equation}
so the true density of the cluster is related to the apparent Newtonian density$\rho_N$ as
\begin{equation}
\rho\left(r\right) = \left({r_0 \over r}\right) \rho_N\left(r\right).
\end{equation}
In other words, assuming Newtonian gravity causes us to {\em underestimate} the baryon fraction in a cluster at a radius $r$ by a factor of $\left(r / r_0\right)$. Taking the characteristic radius to be a typical cluster core radius $r_c \sim 0.2 h^{-1}\,{\rm Mpc}$,$^5$ we can relate the apparent baryon fraction $f^{N}_{\rm B}$ to the true baryon fraction as
\begin{equation}
f_{\rm B} \equiv 1 = \left({r_c \over r_0}\right) f^N_{\rm B}.
\end{equation}
We then have an estimate for the scale $r_0$ that agrees amazingly well with what we would expect from galactic dynamics,
\begin{eqnarray}
r_0 =&& 1.2 h^{-5/2}\,{\rm kpc}\cr
\simeq&& 4.3\,{\rm kpc}\ \ \left(h = 0.6\right).
\end{eqnarray}
Our assumption of a baryon-only universe and an extra force works better than we perhaps have any right to expect, explaining (at least roughly) the independent phenomena of galactic disk dynamics and galaxy cluster dynamics within a single simple framework.

Unfortunately, our run of luck ends when we consider gravitational lensing by clusters. Since our hypothetical force couples to baryon number as a charge, we expect it to interact with photons only via loop effects. Therefore gravitational lensing by the cluster will measure only the gravitational potential, or, equivalently, the actual mass of the cluster. We therefore expect lensing mass estimates to be significantly lower than dynamical mass estimates, ``underestimating'' the cluster mass by a factor of $1 / f^N_{\rm B} \sim 10$. In fact, this is not so. In a survey of lensing mass estimates of clusters, Wu {\it et al.} find that weak lensing mass estimates agree well with both mass estimates determined from velocity dispersion and from hydrostatic equilibrium.$^6$ Wu {\it et al.} in fact find that mass estimates from strong lensing tend to {\em overestimate} the the cluster masses by a significant amount. There is no evidence for systematically low mass estimates that would be required if our extra force model were correct. Lensing kills the model.

\section{Conclusions}

We have presented a hypothesis for a way to construct a universe with no dark matter, in which an extra force with coupling to baryon number dominates over gravitation on scales larger than a few kpc. A logarithmic potential results in naturally flat galactic rotation curves and very consistently explains the apparent baryon fraction of galaxy clusters $f_{\rm B} < 1$ as an artifact of the new force. Gravitational lensing, however, can't be fooled, since it measures the actual mass of galaxy clusters, regardless of the force law responsible for their dynamics. The model fails. 

So why bother even talking about such an alternative model? First, it brings to light a general fact about alternatives to dark matter. Whatever our ``extra'' force might be, it must couple to photons (lensing) in exactly the same way it couples to ordinary matter (dynamics). In other words, it must act like gravity. And anything that is coupled to matter and radiation exactly like gravity must, in fact, {\em be} gravity. This is suggestive of the conclusion that there is simply no way to do away with dark matter without significant modifications to General Relativity itself. 

Finally, although our attempt to invoke a new force as a way to eliminate dark matter ultimately fails, such additional forces could still be of cosmological interest. A particularly interesting question is what effect a weak long-range force would have on the evolution of the universe as a whole, independent of any assumptions of the composition of the matter in the universe.

\end{document}